Luke Peters[1,2]*, Davide Rocco[3], Luana Olivieri[1,2], Unai Arregui Leon[4], Vittorio Cecconi[1], Luca Carletti[3], Carlo Gigli[5], Giuseppe Della Valle[4], Antonio Cutrona [1,2], Juan Sebastian Totero Gongora[1,2], Giuseppe Leo[5], Alessia Pasquazi[1,2], Costantino De Angelis[3], Marco Peccianti[1,2]*.

1. Emergent Photonics Research Centre, Dept. of Physics, Loughborough University, Loughborough, LE11 3TU, England, UK

2. Emergent Photonics Lab (Epic), Department of Physics and Astronomy, University of Sussex, Brighton, BN1 9QH, UK

3. University of Brescia, Department of Information Engineering, via Branze 38, 25123, Brescia, Italy

4. Politecnico di Milano, Department of Physics, Piazza Leonardo Da Vinci 32, 20133, Milan, Italy

5. University of Paris, Matériaux et Phénomènes Quantiques, 10 rue A. Domon et L. Duquet, 75013, Paris, France

**Abstract**

Metasurfaces represent a new frontier in materials science paving for unprecedented methods of controlling electromagnetic waves, with a range of applications spanning from sensing to imaging and communications. For pulsed terahertz generation, metasurfaces offer a gateway to tuneable thin emitters that can be utilised for large-area imaging, microscopy and spectroscopy. In literature THz-emitting metasurfaces generally exhibit high absorption, being based either on metals or on semiconductors excited in highly resonant regimes. Here we propose the use of a fully dielectric semiconductor exploiting morphology-mediated resonances and inherent quadratic nonlinear response. Our system exhibits a remarkable 40-fold efficiency enhancement compared to the unpatterned at the peak of the optimised wavelength range, demonstrating its potential as scalable emitter design.

1. **Introduction**

Optical metasurfaces, the two-dimensional counterpart of metamaterials, have emerged as a compelling area of research in photonics. In general terms, metasurfaces are densely engineered surfaces and are often composed of arrays of subwavelength structures where the field is manipulated by different physical local processes. Research has been motivated by the promise of manipulating light in novel fashions, thereby opening up new avenues for optical functionalities [1]. The studies in this particular area are driven by the aspiration to develop metasurface-based devices that can ensure efficient, compact, and on-chip frequency conversion, steering the future of photonics towards a more integrated and functional landscape, with very strong use-cases beginning to enter the mainstream, ranging from flat-optics cameras on mobile phones [2] to wave-front engineering for telecommunications [3]. Within the framework of the nonlinear frequency conversion, i.e the generation of new optical frequencies via a nonlinear field-matter interaction, metasurfaces are assuming a quite intriguing pivotal role, as they offer a unique platform to enhance the nonlinear optical processes, overcoming the limitations posed by conventional nonlinear materials such as phase-matching constraints and low conversion efficiencies [4–6]. The investigation of this concepts within the terahertz frequency spectrum is quite recent. General research on terahertz (THz) light, which occupies the 0.1 to 10 THz band of the electromagnetic spectrum, is fuelled by fundamental and practical ramifications in fields as different as biological imaging [7–9] art restoration [10–12] and telecommunications [13–15]. Indeed, nonlinear generation of terahertz pulses from ultrafast optical sources is indeed standard in the field [16–18] and it is historically considered a seminal achievement enabling the modern terahertz research area. However, the lack of efficient large-area thin emitters is certainly a fundamental and practical limit, bringing cumbersome experimental setups with complex geometries as well as several other limitations like the resolution limits of novel near-field imaging systems [19–23]. Research on ultrathin emitters brought the investigation of surfaces because of the rich nonlinear optical phenomenology hosted [24–27] and also to the exploration of novel physical mechanism as in the emerging class of spintronic emitters [28–30] with exceptional conversion efficiency per unit of thickness. However, in those technologies the ability of tailoring the emission in regard to the the nature of the excitation and the spatio-temporal property of the emitted waves are quite limited. As a result, we expect metasurfaces to play a growing role in this area [31–33].

Among the platforms proposed in the community, metal-based metasurfaces [34–36] offers a relatively strong nonlinear response boosted by local field enhancement without the several challenges related to the typical velocity matching condition required in bulk generation. Metal high ohmic losses at optical frequencies are the typical drawback that affects nonlinear conversion efficiencies [37,38].

More exotic metasurfaces platform includes: randomly distributed nanotips such as "black-silicon" [39,40], and "quasi-Bound in Continuum (BIC)" materials [41]. Although these results demonstrate enhanced THz generation, they seldom offer a means of customizing emission parameters, such as phase, wavelength, bandwidth. In addition, quasi-BIC exploits narrowband resonances which also might affect typical terahertz use-cases. Recently, terahertz generation from above-bandgap excited semiconductor metasurfaces has been demonstrated [42] in absorption regime. The generation exceeds the one from the native substrate, usually driven by a combination of different surface phenomena, e.g. surface quadratic nonlinearities, carrier drift and photo-Dember processes [43] and the bulk nonlinear response.

Drawing insights from the field of Mie resonances, which employs dielectrics to mitigate conversion losses [44–46], here we present dielectric metasurfaces tailored for pulsed terahertz generation. Our system is based on popular AlGaAs/GaAs semiconductor platform, hence enabling agile integration in photonics chips. The dielectric metasurfaces showcased herein exhibit a local efficiency that is approximately 40-fold the conversion from the blank substrate at resonance.

## 2. Methodology and results

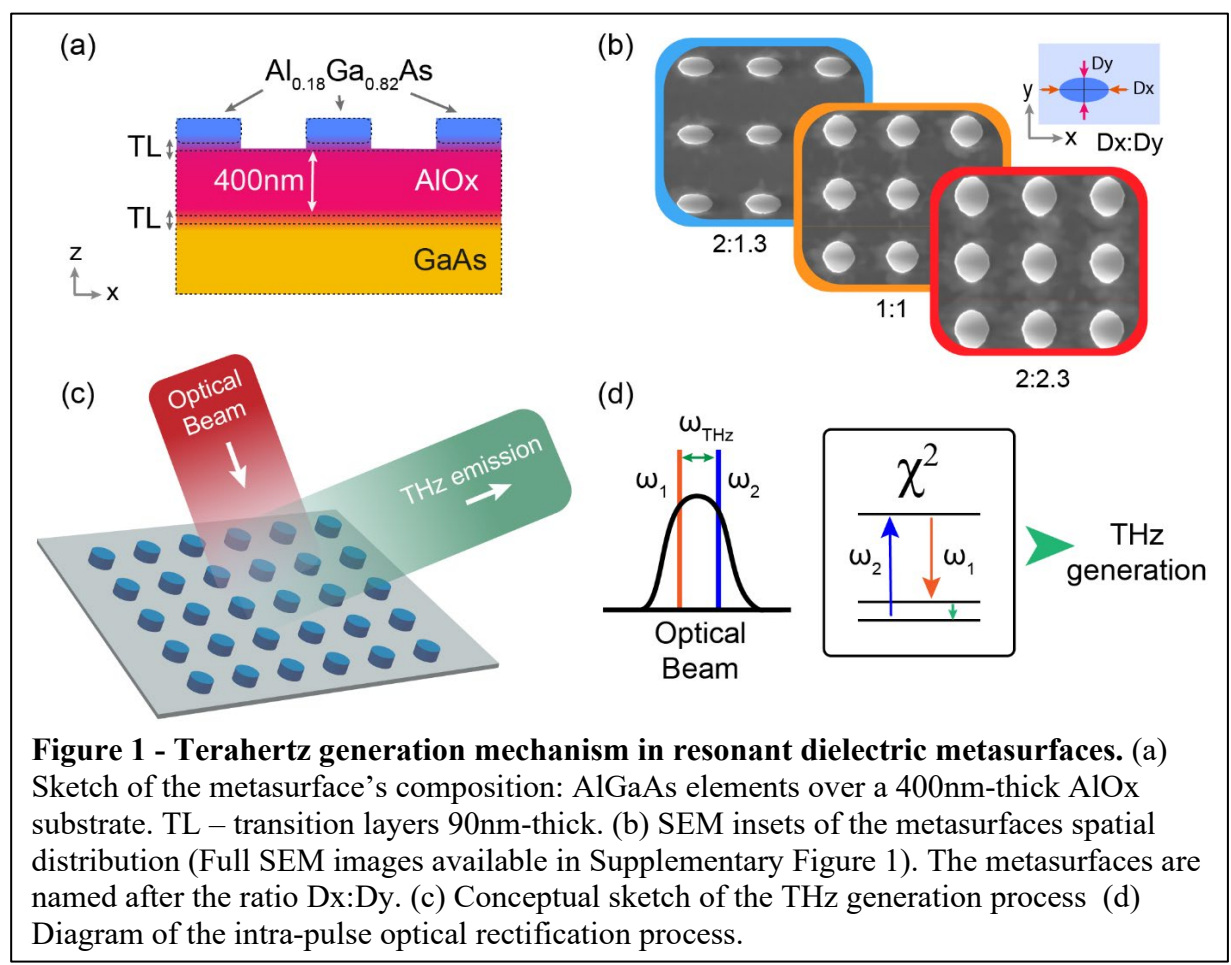

**Figure 1 - Terahertz generation mechanism in resonant dielectric metasurfaces.** (a) Sketch of the metasurface's composition: AlGaAs elements over a 400nm-thick AlOx substrate. TL – transition layers 90nm-thick. (b) SEM insets of the metasurfaces spatial distribution (Full SEM images available in Supplementary Figure 1). The metasurfaces are named after the ratio Dx:Dy. (c) Conceptual sketch of the THz generation process (d) Diagram of the intra-pulse optical rectification process.

The metasurfaces comprise 400nm tall nanocylinder of $Al_{0.18}Ga_{0.82}As$ fabricated using electron-beam lithography. These cylinders are placed on a 400nm-thick AlOx substrate with a low refractive index (n=1.6) (Fig. 1a). The fabrication process is detailed in the methods section. We examined three different designs characterized by a bidimensional array of nanocylinders with an elliptical cross section, each with distinct ellipticities Dx:Dy, where Dx and Dy are the axis of the elliptical base Fig. 1b displays an inset for the 2:1.3, 1:1, and 2:2.3 ellipticity metasurfaces.

Specifically, the 1:1 design has a circular base with a 190nm radius, 2:1.3 has dimensions of 384nm by 207nm, and 2:2.3 measures 380nm by 437nm. The period of the bidimensional array is 750nm. Comprehensive SEM images can be found in the supplementary section (figure S1).

The samples are analyzed using a standard reflection geometry, employing a high-energy optical beam for the nonlinear generation of THz radiation as schematically sketched in (Fig. 1c). The structures are tailored to maximize internal energy for the optical input wavelength, approximately 1280nm (below-bandgap), and for a p-polarized coupling at 45 degrees optical impinging angle.

As the structure offers negligible absorption (unlike for example typical surface generation in low-bandgap III-V semiconductors), the generation is mediated by a quadratic field-matter interaction mechanism that drives an optical rectification process, as in many THz nonlinear crystals [47] (Fig. 1d).

3D Finite Element simulations of the metasurfaces were used to correlate the different ellipticities with the spectrum of the internal energy (see Fig.2a-d) (more information in methods), highlighting a predominant resonance around the pump wavelength (1280nm) for the sample 2:2.3.

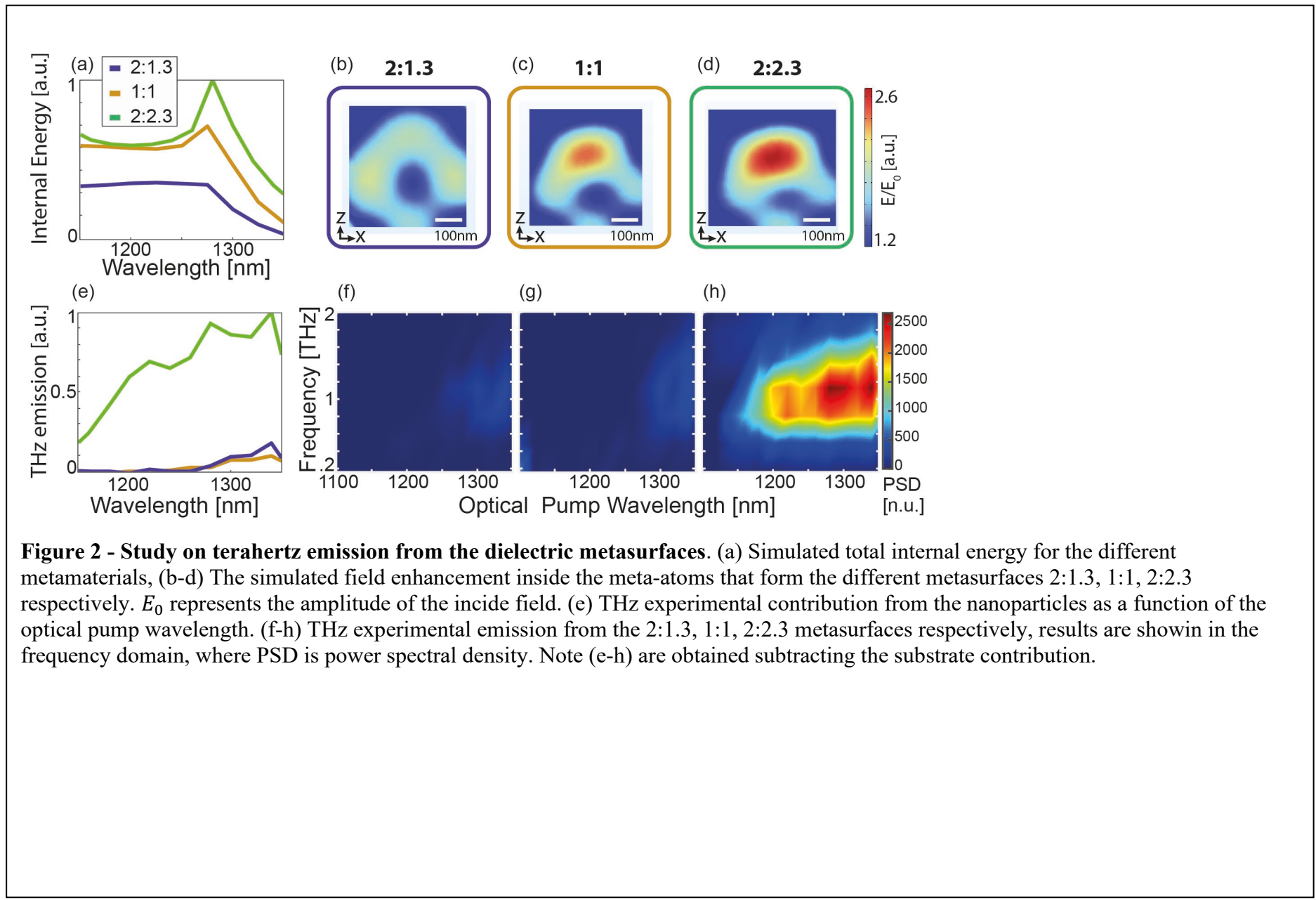

**Figure 2 - Study on terahertz emission from the dielectric metasurfaces**. (a) Simulated total internal energy for the different metamaterials, (b-d) The simulated field enhancement inside the meta-atoms that form the different metasurfaces 2:1.3, 1:1, 2:2.3 respectively. $E_0$ represents the amplitude of the incide field. (e) THz experimental contribution from the nanoparticles as a function of the optical pump wavelength. (f-h) THz experimental emission from the 2:1.3, 1:1, 2:2.3 metasurfaces respectively, results are showin in the frequency domain, where PSD is power spectral density. Note (e-h) are obtained subtracting the substrate contribution.

Terahertz emission efficiency was tested via focussed ultrafast femtosecond pulse illumination, with a spot diameter in the order of 50 μm. Fig.2f-h presents the general feature of the emission at different pump centre wavelengths and for different aspect ratios of the structure, in the range 1100-1350 nm. Notably the contribution of the substrate to emission was also measured (reported in supplementary Figure S2 for completeness) and removed from the characterisation of the metasurfaces.

The theoretical modelling predicts an enhancement of the total internal energy for nanoparticles with a longer axis parallel to the terahertz polarization direction (2:2.3 design) around the optical pump wavelength of 1280nm (Fig. 2a,d). This alignment allows the optical field to couple more effectively to dielectric resonances along the longer nanoparticle axis, boosting the internal field intensity. A larger internal field enables greater nonlinear optical rectification, leading to higher terahertz emission efficiency. Notably, beyond 1280nm, the experimental results deviate from the modelling predictions (Fig 2a,e). This discrepancy arises most likely because the simulations account only for nonlinear generation within the metasurface nanoparticles, neglecting substrate contributions.

The experiments clearly evidence a significant boost in terahertz emission from the 2:2.3 metasurfaces at the 1280nm (Fig. 2e,h). Across the full span of pump wavelengths tested, this elongated nanoparticle design maintains higher conversion efficiency, validating the enhanced performance of this metasurface configuration. This is additionally confirmed by the nonlinear simulations (reported in supplementary Figure S3). Remarkably, a 40-fold increase in the total detected energy is observed when comparing metasurface 2:2.3 (Fig 2h) to the base substrate (Fig S3a) at a pump wavelength of 1400n. For the designed resonance peak at 1280nm an enhancement of around 15 times is seen.

Although the other two metasurface geometries produce lower terahertz yields, they reveal quite intriguing emission dynamics. Terahertz time-domain spectra for each design at various optical wavelengths (Fig. 3a-c) show a distinct change of the carrier envelope phase of the pulse emitted. In a further investigation, we plot the terahertz phase for the different metasurfaces across a range of pump wavelengths and terahertz frequencies (Fig. 3d-f). Remarkably, these broadband resonant nanostructures enable continuous control over the emitted terahertz phase profile simply by acting upon the metasurface shape and pump wavelength. We remark that this could allow a route towards wavefront engineering, which is a sharp challenge in broadband terahertz domain.

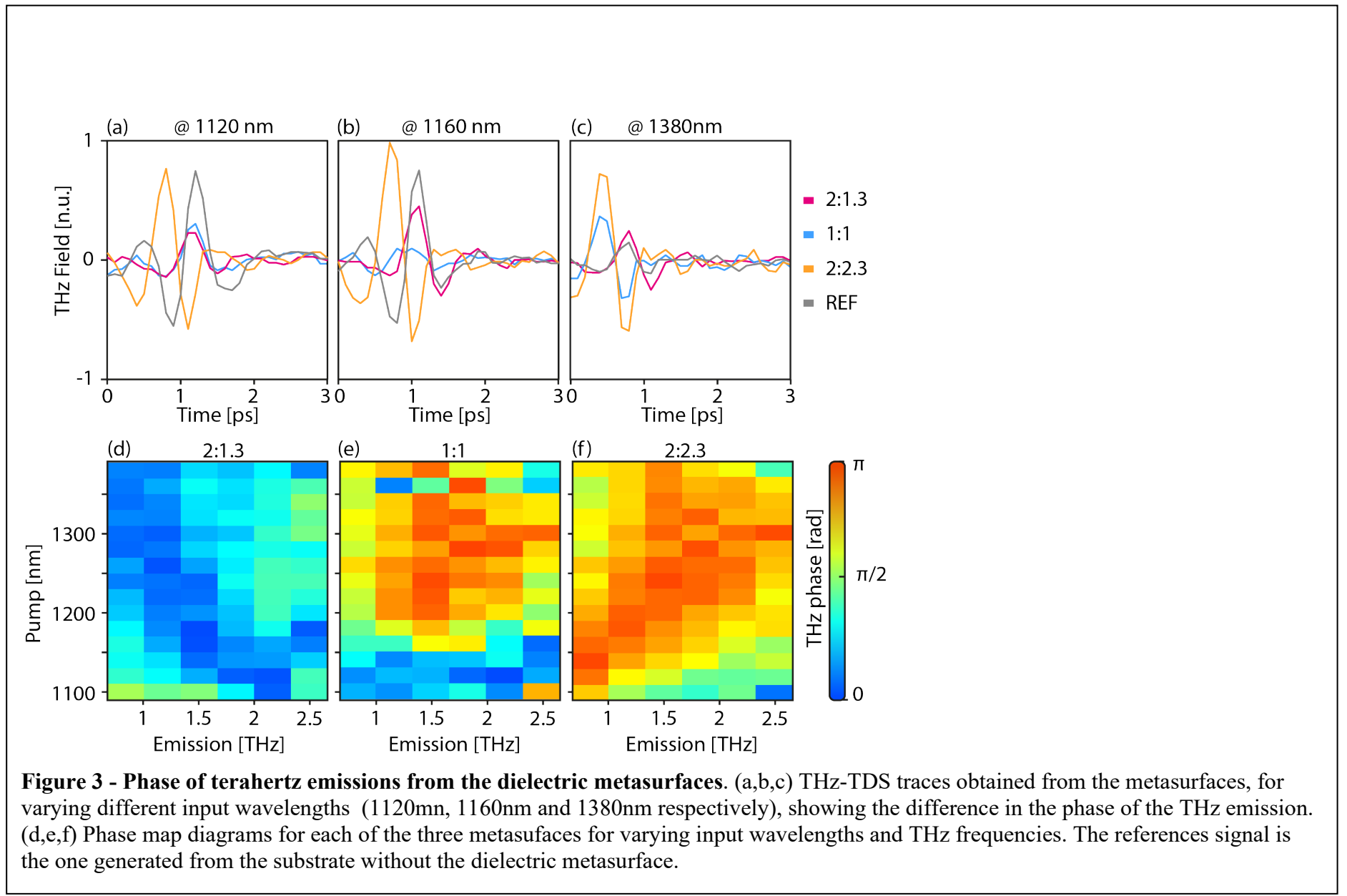

**Figure 3 - Phase of terahertz emissions from the dielectric metasurfaces**. (a,b,c) THz-TDS traces obtained from the metasurfaces, for varying different input wavelengths (1120mn, 1160nm and 1380nm respectively), showing the difference in the phase of the THz emission. (d,e,f) Phase map diagrams for each of the three metasufaces for varying input wavelengths and THz frequencies. The references signal is the one generated from the substrate without the dielectric metasurface.

## 3. Conclusions

In conclusion, this work demonstrates broadband all-dielectric metasurfaces for efficient, customizable terahertz pulse generation. The proposed structure comprises cylindrical semiconductor nano-resonators based on the AlGaAs/GaAs platform. Through optimisation of the metasurface design, a significant 40-fold enhancement in terahertz emission efficiency is achieved compared to the bare substrate. Intriguing, beyond the efficiency, the structure enables form of control over the emitted terahertz radiation characteristics. By tailoring the shape and dimensions of the nano-cylinder elliptical cross-section, the amplitude and the phase profile of the resulting terahertz pulses can be changed. The structures enable manipulation of the terahertz phase by merely tuning the optical excitation wavelength. This peculiar capacity for spatio-temporal structuring of the terahertz wavefront at a planar nanophotonic interface could empower different type applications. These include ultrafast terahertz beam steering, wireless communication, computational imaging, and spectroscopy. More broadly, the approach highlights the potential of all-dielectric metasurfaces for nonlinear frequency conversion, circumventing limitations of plasmonic systems and, more in general, non-transparent systems. With further optimization, such engineered nano-resonator surfaces may displace traditional bulk crystals for terahertz generation. Looking beyond, the concepts demonstrated here is a potential route to unlock the full potential of metasurfaces in terahertz photonics.

## 4. Methods

### 4.1. Fabrication

Sample fabrication is based in a molecular beam epitaxy growth substrate with the following layered structure: (i) [100] non-intentionally doped GaAs wafer, (ii) 90 nm of GaAs-to-Al0.98Ga0.02As transition layer with linearly increasing Al concentration, (iii) 1 μm of Al0.98Ga0.02As, 90 nm of Al0.98Ga0.02As-to-Al0.18Ga0.82As transition layer with linearly decreasing Al concentration, (iv) 400 nm of Al0.18Ga0.82As For the adhesion layer, we deposited a 10 nm layer of SiO2 using plasma-enhanced chemical vapor deposition (PECVD). This was complemented by a spin-coated 100 nm layer of hydrogen silsesquioxane (HSQ) serving as a negative resist. The desired metasurface geometry was patterned through 20 kV e-beam lithography and transferred to the AlGaAs layer by inductively coupled plasma-reactive-ion etching (ICP- RIE) with SiCl4:Ar chemical treatment. Finally, we performed wet oxidation in an oven at 390 °C for 40 minutes under a controlled mix of water vapor and N2:H2 gas carrier aiming at converting the 1 μm Al-rich layer in amorphous non stochiometric aluminum oxide (AlOx).

### 4.2. Experimental Setup

Our experimental setup consists of a Chameleon ultrafast laser, equipped with a compact OPO (optical parametric oscillator) attachment. The laser offers a repetition rate of 80 MHz, a pulse duration of 140 fs before the OPO, and an average power of 4 W before the OPO. The OPO allowed for the variability in the wavelength. This laser fed directed a conventional THz-TDS setup, configured for reflection-generation spectroscopy. Notably, the detection crystal employed was a 1mm-thick Zinc Telluride. The ZnTe was rotated to selectively detect p-polarised input light (with polarization lying in the plane containing impinging and reflected rays) [48]. The detection was performed via optical sampling supplied directly with the chameleon laser line at a wavelength 850 nm (close to optimal collinear phase-matching condition in ZnTe). Diagrams detailing the experimental setting are available in the supplementary material (supplementary Figure S4).

**4.3 Theoretical Simulations**

We perform 3D Finite Element Method simulations in COMSOL Multiphysics. We model the metasurfaces of AlGaAs nanoresonators on a substrate, applying Floquet boundary conditions to emulate an infinite periodic structure. The refractive index dispersion of AlGaAs is taken into account according to [49].

The primary excitation was modelled as a plane wave p-polarized with an incident angle of 45°. We perform both linear and nonlinear calculations. Due to the significant difference in length scales between the optical (a few μm) and THz domains (hundreds to thousands of μm), our simulation incorporated two components with distinct domain dimensions. One component ensures a well-meshed model in the optical domain, featuring a refined mesh for the nano-cylindrical resonators. In contrast, the other component accommodates at least one wavelength in the THz regime.

As mentioned in [49], we implemented a General Extrusion coupling operator to correlate the nanocylinders of both components. This process maps electromagnetic fields in the optical wavelengths in the destination geometry (THz range) by keeping the same finer mesh inside the resonators and relaxing the mesh in the surrounding domains - effectively optimizing computational resource utilization.

For the incident beam, our model is based on a Gaussian pulse centred at 2.34 THz (~1280 nm) with bandwidth 2 THz, defined as the spectral width at $1/e^2$ of the spectral peak. To assess the complete THz emission from our metasurface, we employed a discretized spectrum of the incident Gaussian beam. It is sampled with a discrete set of frequency steps (N=18, each step approximately 0.118 THz apart), enabling the calculation of difference-frequency generation (DFG) between the weighted spectral components. This approach condenses the simulations of the total THz generation into several DFG problems which can be modelled following ref. [49], where only the bulk nonlinearities are considered. For given incident fields spectral components, $[E(\omega_1), E(\omega_2)]$ with their relative amplitude weight, we use Eq. 1 for obtaining the DFG nonlinear current densities $J_i$ at $\omega_n = \omega_2 - \omega_1$.

$$J_i(\omega_n) = -i\omega_n\varepsilon_0\chi^{(2)}_{ijk}(\omega_n;\omega_2,-\omega_1)[E_j(\omega_2)E_k^*(\omega_1) + E_k(\omega_2)E_j^*(\omega_1)] \qquad (1)$$

In the previous equation, $\varepsilon_0$ is the vacuum permittivity and $\chi^{(2)}$ denotes the frequency dependent AlGaAs nonlinear susceptibility tensor, i≠j≠k being the Cartesian coordinates. The nonlinear currents imposed within the AlGaAs nanocylinder are the sources for the nonlinear DFG field. For every spectral component (with N frequency components of the incident Gaussian beam), we calculate the DFG contributions for all combinations, resulting in N(N–1)/2 simulations.

Lastly, the final THz signal component at $\omega_n$ is computed as the sum of all the DFG contributions which generate the nonlinear THz signal at that specific $\omega_n$ (where $\omega_n$ = n·Δf with n=1,…,N–1). The obtained results are in substantial agreement with the experimental measurement. Hence, the simplified computational framework proposed here is able to catch all the main features concerning the THz emission from dielectric metasurfaces. It's important to emphasize that only bulk nonlinearities are considered in our simulations. As shown theoretically in ref. [49] the THz efficiency notably improves around the AlGaAs phonon polariton frequencies (~8 THz, ~ 11 THz). Consequently, leveraging incident pump signals with an expansive frequency band could offer additional advantages in enhancing nonlinear efficiency.

## 5. References

1. N. Yu, P. Genevet, M. A. Kats, F. Aieta, J.-P. Tetienne, F. Capasso, and Z. Gaburro, "Light Propagation with Phase Discontinuities: Generalized Laws of Reflection and Refraction," Science **334**(6054), 333–337 (2011).
2. O. Reshef, M. P. DelMastro, K. K. M. Bearne, A. H. Alhulaymi, L. Giner, R. W. Boyd, and J. S. Lundeen, "An optic to replace space and its application towards ultra-thin imaging systems," Nat. Commun. **12**(1), 3512 (2021).
3. Z. Wang, T. Li, A. Soman, D. Mao, T. Kananen, and T. Gu, "On-chip wavefront shaping with dielectric metasurface," Nat. Commun. **10**(1), 3547 (2019).
4. A. V. Kildishev, A. Boltasseva, and V. M. Shalaev, "Planar Photonics with Metasurfaces," Science **339**(6125), 1232009 (2013).
5. S. Liu, M. B. Sinclair, S. Saravi, G. A. Keeler, Y. Yang, J. Reno, G. M. Peake, F. Setzpfandt, I. Staude, T. Pertsch, and I. Brener, "Resonantly Enhanced Second-Harmonic Generation Using III–V Semiconductor All-Dielectric Metasurfaces," Nano Lett. **16**(9), 5426–5432 (2016).
6. A. Di Francescantonio, A. Zilli, D. Rocco, L. Coudrat, F. Conti, P. Biagioni, L. Duò, A. Lemaître, C. De Angelis, G. Leo, M. Finazzi, and M. Celebrano, "All-optical free-space routing of upconverted light by metasurfaces via nonlinear interferometry," Nat. Nanotechnol. 1–8 (2023).
7. Q. Sun, Y. He, K. Liu, S. Fan, E. P. J. Parrott, and E. Pickwell-MacPherson, "Recent advances in terahertz technology for biomedical applications," Quant. Imaging Med. Surg. **7**(3), 34555–34355 (2017).
8. E. Pickwell, B. E. Cole, A. J. Fitzgerald, M. Pepper, and V. P. Wallace, "*In vivo* study of human skin using pulsed terahertz radiation," Phys. Med. Biol. **49**(9), 1595–1607 (2004).
9. P. C. Ashworth, E. Pickwell-MacPherson, E. Provenzano, S. E. Pinder, A. D. Purushotham, M. Pepper, and V. P. Wallace, "Terahertz pulsed spectroscopy of freshly excised human breast cancer," Opt. Express **17**(15), 12444–12454 (2009).
10. C. Seco-Martorell, V. López-Domínguez, G. Arauz-Garofalo, A. Redo-Sanchez, J. Palacios, and J. Tejada, "Goya's artwork imaging with Terahertz waves," Opt. Express **21**(15), 17800 (2013).
11. L. Öhrström, B. M. Fischer, A. Bitzer, J. Wallauer, M. Walther, and F. Rühli, "Terahertz imaging modalities of ancient Egyptian mummified objects and of a naturally mummified rat," Anat. Rec. Hoboken NJ 2007 **298**(6), 1135–1143 (2015).

12. G. Leong, M. Brolly, P. Taday, and D. Giovannacci, "Towards terahertz imaging applications at Stonehenge for identification of prehistoric carvings," in *2022 47th International Conference on Infrared, Millimeter and Terahertz Waves (IRMMW-THz)* (2022), pp. 1–2.
13. T. Nagatsuma, G. Ducournau, and C. C. Renaud, "Advances in terahertz communications accelerated by photonics," Nat. Photonics **10**(6), 371–379 (2016).
14. H. Elayan, O. Amin, B. Shihada, R. M. Shubair, and M.-S. Alouini, "Terahertz Band: The Last Piece of RF Spectrum Puzzle for Communication Systems," IEEE Open J. Commun. Soc. **1**, 1–32 (2020).
15. A. E. Willner, X. Su, H. Zhou, A. Minoofar, Z. Zhao, R. Zhang, M. Tur, A. F. Molisch, D. Lee, and A. Almaiman, "High capacity terahertz communication systems based on multiple orbital-angular-momentum beams," J. Opt. **24**(12), 124002 (2022).
16. K. H. Yang, P. L. Richards, and Y. R. Shen, "Generation of Far-Infrared Radiation by Picosecond Light Pulses in $LiNbO_3$," Appl. Phys. Lett. **19**(9), 320–323 (1971).
17. G. Mourou, C. V. Stancampiano, A. Antonetti, and A. Orszag, "Picosecond microwave pulses generated with a subpicosecond laser-driven semiconductor switch," Appl. Phys. Lett. **39**(4), 295–296 (1981).
18. D. H. Auston, K. P. Cheung, J. A. Valdmanis, and D. A. Kleinman, "Cherenkov Radiation from Femtosecond Optical Pulses in Electro-Optic Media," Phys. Rev. Lett. **53**(16), 1555–1558 (1984).
19. F. Blanchard, A. Doi, T. Tanaka, H. Hirori, H. Tanaka, Y. Kadoya, and K. Tanaka, "Real-time terahertz near-field microscope," Opt. Express **19**(9), 8277–8284 (2011).
20. J. S. Totero Gongora, L. Olivieri, L. Peters, J. Tunesi, V. Cecconi, A. Cutrona, R. Tucker, V. Kumar, A. Pasquazi, and M. Peccianti, "Route to Intelligent Imaging Reconstruction via Terahertz Nonlinear Ghost Imaging," Micromachines **11**(5), 521 (2020).
21. L. Olivieri, J. S. T. Gongora, L. Peters, V. Cecconi, A. Cutrona, J. Tunesi, R. Tucker, A. Pasquazi, and M. Peccianti, "Hyperspectral terahertz microscopy via nonlinear ghost imaging," Optica **7**(2), 186 (2020).
22. L. Olivieri, L. Peters, V. Cecconi, A. Cutrona, M. Rowley, J. S. Totero Gongora, A. Pasquazi, and M. Peccianti, "Terahertz Nonlinear Ghost Imaging via Plane Decomposition: Toward Near-Field Micro-Volumetry," ACS Photonics (2023).
23. L. L. Hale, T. Siday, and O. Mitrofanov, "Near-field imaging and spectroscopy of terahertz resonators and metasurfaces [Invited]," Opt. Mater. Express **13**(11), 3068–3086 (2023).
24. P. Gu, M. Tani, S. Kono, K. Sakai, and X.-C. Zhang, "Study of terahertz radiation from InAs and InSb," J. Appl. Phys. **91**(9), 5533–5537 (2002).
25. M. Reid and R. Fedosejevs, "Terahertz emission from (100) InAs surfaces at high excitation fluences," Appl. Phys. Lett. **86**(1), 011906–011906 (2005).
26. A. Krotkus, R. Adomavičius, G. Molis, and V. L. Malevich, "TeraHertz Emission from InAs Surfaces Excited by Femtosecond Laser Pulses," J. Nanoelectron. Optoelectron. **2**(1), 108–114 (2007).
27. L. Peters, J. Tunesi, A. Pasquazi, and M. Peccianti, "High-energy terahertz surface optical rectification," Nano Energy **46**, 128–132 (2018).
28. T. Kampfrath, M. Battiato, P. Maldonado, G. Eilers, J. Nötzold, S. Mährlein, V. Zbarsky, F. Freimuth, Y. Mokrousov, S. Blügel, M. Wolf, I. Radu, P. M. Oppeneer, and M. Münzenberg, "Terahertz spin current pulses controlled by magnetic heterostructures," Nat. Nanotechnol. **8**(4), 256–260 (2013).
29. T. Seifert, S. Jaiswal, U. Martens, J. Hannegan, L. Braun, P. Maldonado, F. Freimuth, A. Kronenberg, J. Henrizi, I. Radu, E. Beaurepaire, Y. Mokrousov, P. M. Oppeneer, M. Jourdan, G. Jakob, D. Turchinovich, L. M. Hayden, M. Wolf, M. Münzenberg, M. Kläui, and T. Kampfrath, "Efficient metallic spintronic emitters of ultrabroadband terahertz radiation," Nat. Photonics **10**(7), 483–488 (2016).
30. C. Bull, S. M. Hewett, R. Ji, C.-H. Lin, T. Thomson, D. M. Graham, and P. W. Nutter, "Spintronic terahertz emitters: Status and prospects from a materials perspective," APL Mater. **9**(9), 090701 (2021).
31. M. Tal, S. Keren-Zur, and T. Ellenbogen, "Nonlinear Plasmonic Metasurface Terahertz Emitters for Compact Terahertz Spectroscopy Systems," ACS Photonics **7**(12), 3286–3290 (2020).
32. S. Keren-Zur, O. Avayu, L. Michaeli, and T. Ellenbogen, "Nonlinear Beam Shaping with Plasmonic Metasurfaces," ACS Photonics **3**(1), 117–123 (2016).
33. S. Keren-Zur, M. Tal, S. Fleischer, D. M. Mittleman, and T. Ellenbogen, "Generation of spatiotemporally tailored terahertz wavepackets by nonlinear metasurfaces," Nat. Commun. **10**(1), 1–6 (2019).
34. H. Cai, Q. Huang, X. Hu, Y. Liu, Z. Fu, Y. Zhao, H. He, and Y. Lu, "All-Optical and Ultrafast Tuning of Terahertz Plasmonic Metasurfaces," Adv. Opt. Mater. **6**(14), 1800143 (2018).
35. J. Lee, M. Tymchenko, C. Argyropoulos, P.-Y. Chen, F. Lu, F. Demmerle, G. Boehm, M.-C. Amann, A. Alù, and M. A. Belkin, "Giant nonlinear response from plasmonic metasurfaces coupled to intersubband transitions," Nature **511**(7507), 65–69 (2014).

36. M. S. Bin-Alam, O. Reshef, Y. Mamchur, M. Z. Alam, G. Carlow, J. Upham, B. T. Sullivan, J.-M. Ménard, M. J. Huttunen, R. W. Boyd, and K. Dolgaleva, "Ultra-high-Q resonances in plasmonic metasurfaces," Nat. Commun. **12**(1), 974 (2021).
37. A. Vora, J. Gwamuri, N. Pala, A. Kulkarni, J. M. Pearce, and D. Ö. Güney, "Exchanging Ohmic Losses in Metamaterial Absorbers with Useful Optical Absorption for Photovoltaics," Sci. Rep. **4**(1), 4901 (2014).
38. J. B. Khurgin, "Replacing noble metals with alternative materials in plasmonics and metamaterials: how good an idea?," Philos. Trans. R. Soc. Math. Phys. Eng. Sci. **375**(2090), 20160068 (2017).
39. J. Tunesi, L. Peters, J. S. Totero Gongora, L. Olivieri, A. Fratalocchi, A. Pasquazi, and M. Peccianti, "Terahertz emission mediated by ultrafast time-varying metasurfaces," Phys. Rev. Res. **3**(4), L042006 (2021).
40. W. Hoyer, A. Knorr, J. V. Moloney, E. M. Wright, M. Kira, and S. W. Koch, "Photoluminescence and terahertz emission from femtosecond laser-induced plasma channels," Phys. Rev. Lett. **94**(11), 115004–115004 (2005).
41. L. Hu, B. Wang, Y. Guo, S. Du, J. Chen, J. Li, C. Gu, and L. Wang, "Quasi-BIC Enhanced Broadband Terahertz Generation in All-Dielectric Metasurface," Adv. Opt. Mater. **10**(12), 2200193 (2022).
42. L. L. Hale, H. Jung, S. D. Gennaro, J. Briscoe, C. T. Harris, T. S. Luk, S. J. Addamane, J. L. Reno, I. Brener, and O. Mitrofanov, "Terahertz Pulse Generation from GaAs Metasurfaces," ACS Photonics 7 (2022).
43. L. Peters, J. Tunesi, A. Pasquazi, and M. Peccianti, "Optical Pump Rectification Emission: Route to Terahertz Free-Standing Surface Potential Diagnostics," Sci. Rep. **7**(1), 9805 (2017).
44. J. S. T. Gongora, G. Favraud, and A. Fratalocchi, "Fundamental and high-order anapoles in all-dielectric metamaterials via Fano–Feshbach modes competition," Nanotechnology **28**(10), 104001 (2017).
45. D. Rocco, V. F. Gili, L. Ghirardini, L. Carletti, I. Favero, A. Locatelli, G. Marino, D. N. Neshev, M. Celebrano, M. Finazzi, G. Leo, and C. D. Angelis, "Tuning the second-harmonic generation in AlGaAs nanodimers via non-radiative state optimization [Invited]," Photonics Res. **6**(5), B6–B12 (2018).
46. V. F. Gili, L. Ghirardini, D. Rocco, G. Marino, I. Favero, I. Roland, G. Pellegrini, L. Duò, M. Finazzi, L. Carletti, A. Locatelli, A. Lemaître, D. Neshev, C. D. Angelis, G. Leo, and M. Celebrano, "Metal–dielectric hybrid nanoantennas for efficient frequency conversion at the anapole mode," Beilstein J. Nanotechnol. **9**(1), 2306–2314 (2018).
47. M. A. A. Rice Y. Jin, X. F. Ma, X. -C. Zhang, D. Bliss, J. Larkin, "Terahertz optical rectification from 〈110〉 zinc-blende crystals," Appl. Phys. Lett. **64**(11), 1324–1326 (1994).
48. P. C. M. Planken, H.-K. Nienhuys, H. J. Bakker, and T. Wenckebach, "Measurement and calculation of the orientation dependence of terahertz pulse detection in ZnTe," J. Opt. Soc. Am. B **18**(3), 313–313 (2001).
49. U. A. Leon, D. Rocco, L. Carletti, M. Peccianti, S. Maci, G. Della Valle, and C. De Angelis, "THz-photonics transceivers by all-dielectric phonon-polariton nonlinear nanoantennas," Sci. Rep. **12**(1), 4590 (2022).

## 6. Funding Acknowledgements

This project received funding from the European Research Council (ERC) under the European Union's Horizon 2020 Research and Innovation Programme grant no. 725046 as well from the UK Engineering and Physical Sciences Research Council (EPSRC), grant no. EP/W028344/1, The Leverhulme Trust Early Career Fellowship Early Career Fellowship (ECF-2022-710 and ECF-2023-315) as well as the Leverhulme Trust research project grant RPG-2022-090.  We acknowledge the Italian Ministry of University and Research (MUR) as part of the PRIN 2022 project GRACE6G (2022H7RR4F), PRIN PNRR 2022 project FLAIRS (P2022RFF9K). This publication is part of the METAFAST project that received funding from the European Union Horizon 2020 Research and Innovation programme under Grant Agreement No. 899673. This work reflects only author views, and the European Commission is not responsible for any use that may be made of the information it contains.

Resonant Fully Dielectric metasurfaces for ultrafast Terahertz pulse generation

*Luke Peters[1,2]\*, Davide Rocco[3], Luana Olivieri[1,2], Unai Arregui Leon[4], Vittorio Cecconi[1], Luca Carletti[3],* Carlo Gigli[5]*, Giuseppe Della Valle[4], Antonio Cutrona[1,2], Juan Sebastian Totero Gongora[1,2], Giuseppe Leo[5], Alessia Pasquazi[1,2], Costantino De Angelis[3], Marco Peccianti[1,2]\**.

1. Emergent Photonics Research Centre, Dept. of Physics, Loughborough University, Loughborough, LE11 3TU, England, UK
2. Emergent Photonics Lab (Epic), Department of Physics and Astronomy, University of Sussex, Brighton, BN1 9QH, UK
3. University of Brescia, Department of Information Engineering, via Branze 38, 25123, Brescia, Italy
4. Politecnico di Milano, Department of Physics, Piazza Leonardo Da Vinci 32, 20133, Milan, Italy
5. University of Paris, Matériaux et Phénomènes Quantiques, 10 rue A. Domon et L. Duquet, 75013, Paris, France

**Supplementary Information**

**List of paper acronyms**

THz - Terahertz

DFG - Difference-frequency generation

SEM - Scanning electron microscope

HSQ - Hydrogen silsesquioxane

PECVD - Plasma-enhanced chemical vapor deposition

ICP-RIE - Inductively coupled plasma-reactive-ion etching

OPO - Optical parametric oscillator

BIC - Bound in Continuum

**Supplementary Figures**

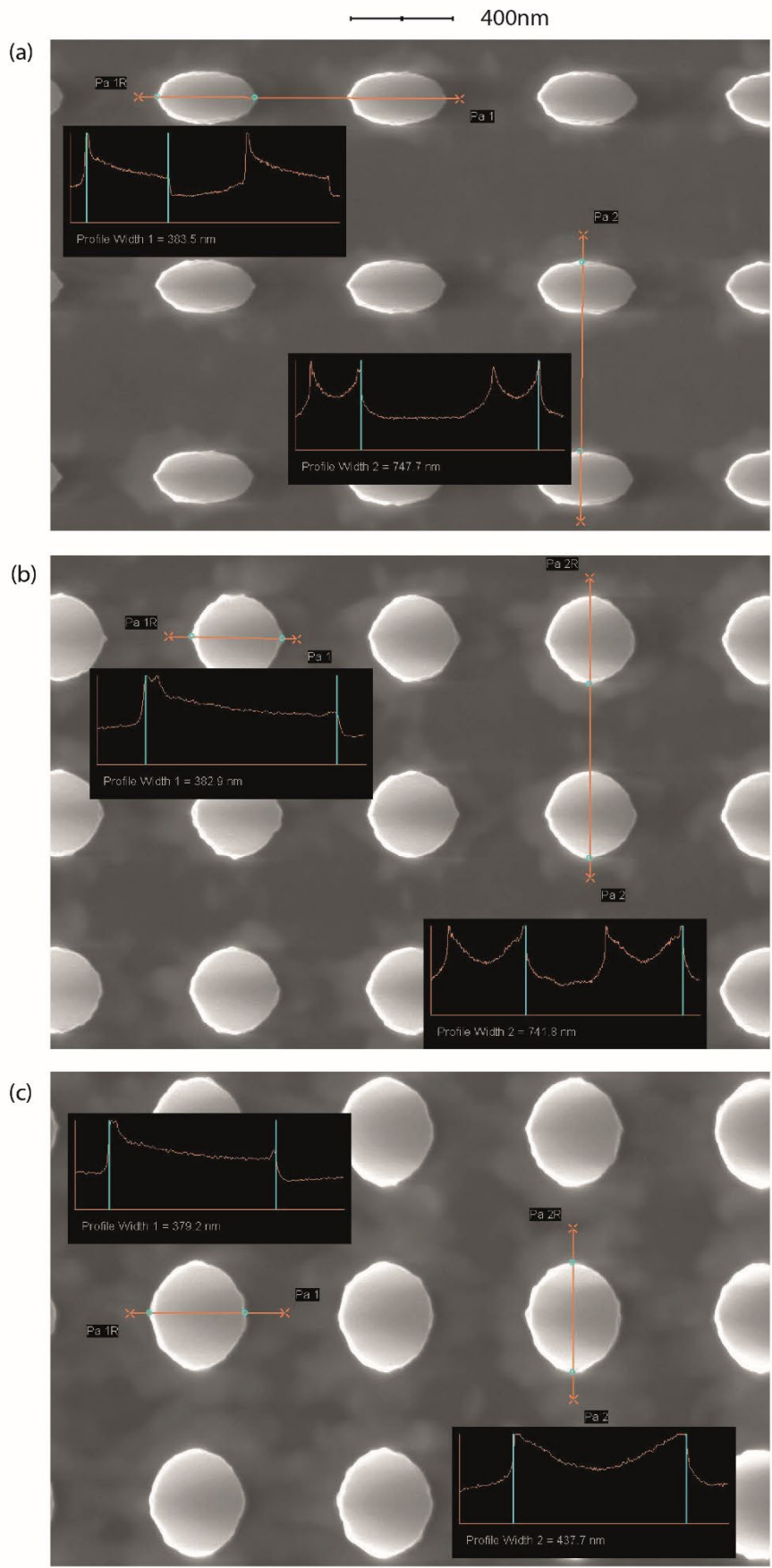

Figure S1: SEM images of the metasurfaces (a) 2:1.3, (b) 1:1, (c) 2:2.3. Showing the spacing between the pillars as well as the diameter of each of the pillars.

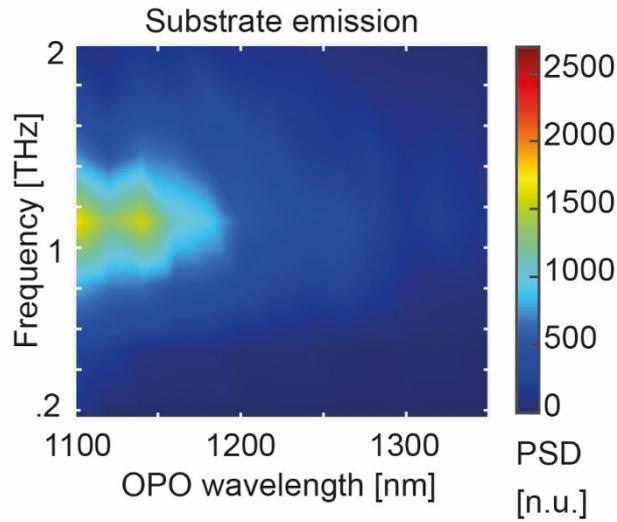

Figure S2: Power Spectral Density (expressed in normalised units) of the substrate contribution to the THz emission at different optical pump wavelengths.

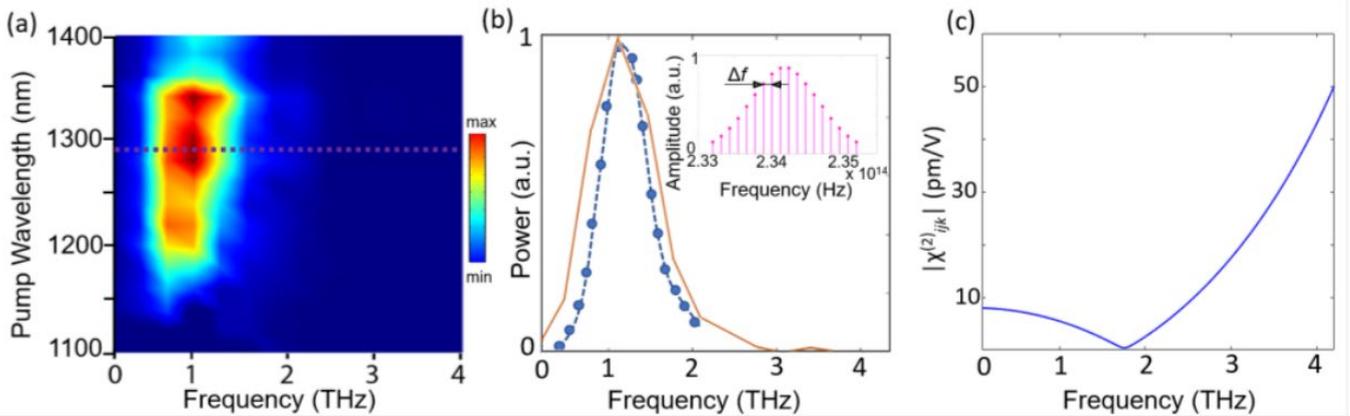

Figure S3: (a) The emitted nonlinear signal intensity as a function of the pump wavelength and the emitted THz frequency for the metasurface 2:2.3. (b) The measured THz signal for an input wavelength of around 1280nm (orange curve, as highlighted by the dashed purple line in (a) and the simulated THz emission for a metasurface with the same geometrical shape as the fabricated one, excited with a gaussian pulse beam centered around 1280nm (blue curve). The inset shows the considered discretized incident spectrum. (c) The amplitude of $\chi^{(2)}_{ijk}$ in the considered frequency range.

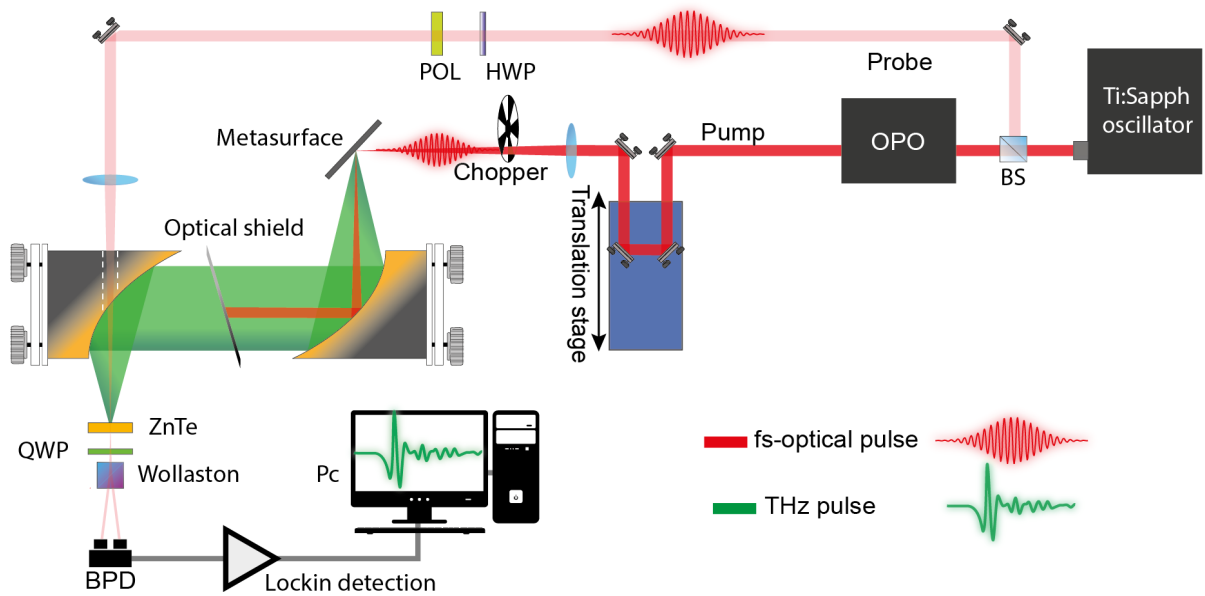

Figure S43: Experimental setup for the THz-TDS. The red and green beam paths denote the OPO and THz beam paths respectively. The THz field is measured with a standard electro-optic detection which retrieves the change of polarisation of an optical probe of energy inside a ZnTe detection crystal due to the THz field. A delay line controls the delay between the THz and the optical probe and allows for the reconstruction of the THz waveform. The THz is generated by the metasurfaces. The optical wavelength is controlled by the tuning of the OPO pump, while the probe wavelength is kept fixed at 800nm.